\documentclass [12pt,twoside]{article}           
\usepackage[left,modulo]{lineno}
\usepackage{setspace}

\countdef\arxiv=201

\ifnum\arxiv=0 \input cpreb.tex \fi

\ifnum\arxiv=1

\usepackage{epsfig,times,lscape}
\usepackage[usenames]{color}

    \ifnum\count102=1

    \fi

    \ifnum\count102=2
\fi

\newcommand{\nc}{\newcommand}

     \ifnum\count101=1
\nc{\qI}[1]{\section{{#1}}}
\nc{\qA}[1]{\subsection{{#1}}}
\nc{\qun}[1]{\subsubsection{{#1}}}
\nc{\qa}[1]{\paragraph{{#1}}}

\def\qpar{\vskip 2mm plus 0.2mm minus 0.2mm}
\def\qL{\hfill \break}
     \fi 

      \ifnum\count101=2
 \nc{\qI}[1]{\parindent=0mm \vskip 8mm 
{\centerline{\LARGE \color{red}#1}}\vskip 3mm}
\nc{\qA}[1]{\vskip 2.5mm \noindent 
{{\bf\large\color{blue}  #1}} \vskip 1mm \parindent=0mm}
 \nc{\qun}[1]{\vskip 1mm \noindent {\sl #1 }\quad }

\def\qL{\hfill \break}
\def\qpar{\vskip 2mm plus 0.2mm minus 0.2mm}

      \fi

\def\qth{\vrule height 12pt depth 0pt width 0pt}
\def\qtb{\vrule height 0pt depth 5pt width 0pt}

\nc{\qfoot}[1]{\footnote{{#1}}}

      \ifnum\count101=1
\def\qbu{\hfill \par \hskip 6mm $ \bullet $ \hskip 2mm}
\def\qee#1{\hfill \par \hskip 6mm (#1) \hskip 2 mm}
      \fi
      \ifnum\count101=2
\def\qbu{\hfill \par \hskip 4mm $ \bullet $ \hskip 2mm}
\def\qee#1{\hfill \par \hskip 4mm (#1) \hskip 2 mm}
      \fi

\def\qparr{ \vskip 1.0mm plus 0.2mm minus 0.2mm \hangindent=10mm
\hangafter=1}

     \ifnum\count101=1 
 
     \fi
     \ifnum\count101=2

  \def\qcitb#1{\noindent \hbox to 102mm{\hfill \small #1} \vskip 1mm}
      \fi

 \def\qpages#1{\count102=0{\loop\advance\count102 by 1
 \null \vfill\eject \ifnum\count102<#1 \repeat}}

\def\qn#1{\eqno \hbox{(#1)}}

\def\qth{\vrule height 12pt depth 0pt width 0pt}
\def\qtb{\vrule height 0pt depth 5pt width 0pt}

\def\qv{\vskip 0.1mm plus 0.05mm minus 0.05mm}
\def\qhu{\hskip 0.6mm}
\def\qhv{\hskip 3mm}

\def\qhw{\hskip 1.5mm}
\def\qleg#1#2#3{\noindent {\bf \small #1\qhw}{\small #2\qhw}{\it \small #3}\qv }
\fi

\begin{document}
\thispagestyle{empty}



\markboth{{\sl \hfill  \hfill \protect\phantom{3}}}
        {{\protect\phantom{3}\sl \hfill  \hfill}}

\color{yellow} 
\hrule height 20mm depth 10mm width 170mm 
\color{black}
\vskip -1.8cm 

 \centerline{\bf \Large Property bubble in Hong Kong:}
\vskip 2mm
 \centerline{\bf \Large A predicted decade-long slump (2016-2025)}
\vskip 10mm
\centerline{\large 
Peter Richmond$ ^1 $ and Bertrand M. Roehner$ ^2 $
}

\vskip 15mm
\large

{\bf Abstract}\quad Between 2003 and 2015
the prices of apartments in Hong Kong (adjusted for inflation)
increased by a factor of 3.8. This is much higher than was observe
in the United States prior to the so-called subprime crisis
of 2007.
The analysis of this 
speculative episode confirms the mechanism and regularities
already highlighted by the present authors in similar episodes
in other countries. Based on these regularities, it
is possible to predict the price trajectory over the 
time interval 2016-2025. It suggests that,
unless appropriate relief is provided by the mainland, 
Hong Kong will experience a decade-long slump. 
Possible implications
for its relations with Beijing
are discussed at the end of the paper.

\vskip 10mm
\centerline{\it Provisional. Version of 3 August 2016. 
Comments are welcome.}
\vskip 10mm

JEL code: D400

{\small Key-words: real estate prices, speculative bubble,
Hong Kong, currency board, price multiplier effect, prediction.}
\vskip 4mm

{\normalsize 
1: School of Physics, Trinity College, Dublin, Ireland.
Email: peter\_richmond@ymail.com \qL
2: Institute for Theoretical and High Energy Physics (LPTHE),
University Pierre and Marie Curie, Paris, France. 
Email: roehner@lpthe.jussieu.fr
}

\vfill\eject

\large

\qI{Introduction}

In addition to being an economic phenomenon, speculative
bubbles have important social aspects. They give rise
to a speculative frenzy which leads buyers
to
forget previous episodes that ended in disasters%
\qfoot{Several instances of speculative frenzy
for various items (stocks, collectible
books, coins and stamps, houses and apartments)
are described (not only qualitatively
but also quantitatively) in Roehner (2001),
particularly in chapter 5 entitled ``Contagion of speculative
frenzy''.}%
.
\qpar

The sheer size of property bubbles ensures they are one of the most important speculative bubbles.
Their main characteristic is that during such episodes
prices are lifted up by the dynamics of speculation. Moreover, in the final stages of the bubble, both
supply and demand
and interest rate levels
become largely irrelevant%
\qfoot{Curiously, many economic studies fail to
take advantage of the great simplification 
permitted by this feature. For instance, a study
entitled ``Is Hong Kong property a bubble?''
(http://discount-investing.com) starts by asking
``Is there a shortage of new housing?''. 
There is of course a shortage of new
housing in all big cities but this question becomes
completely irrelevant during the slump phase of a speculative
episode when the market is dominated
by a flight away from risky assets falling in value.}%
.

It is this aspect which 
allows the effect to be described and predicted by
a fairly simple mechanism as demonstrated 
in previous papers (Roehner 2001, Richmond 2007, 
Richmond et al. 2013).
 In such episodes one can define three prices
which provide a schematic description of the price trajectory:
the price $ p_1 $ at the beginning of the upward phase,
the price $ p_2 $ at the top of the peak and the 
price $ p_3 $ at the end of the downward phase. In relation
with these prices, one can also define the corresponding
amplitudes: $ A_1=p_2/p_1 $ and $ A_2 =p_2/p_3 $.
\qpar

The basic rule which emerges from the study of previous
historical episodes is that $ p_3 $ is only slightly higher
than $ p_1 $. In short, 
the more prices go up, the more they must
come down. Moreover, the duration $ \tau_2 $ of the declining 
phase is only slightly shorter than the duration $ \tau_1 $
of the rising phase.
In other words,
the most basic approximation (a more accurate picture will
be given subsequently)
is that speculative price
peaks are symmetrical with respect to their two
phases%
\qfoot{This pattern has been observed repeatedly, yet
it is still largely overlooked by ``experts''. Thus,
the predictions offered on the Internet 
in the case of Hong Kong rarely cover
more than one year and we did not find any which covers the
next decade.}%
.
\qpar

Needless to say, such predictions cannot
take into account external factors such as government interventions
or major disruptions (e.g. the end of the Soviet Union in 1990
or the worldwide financial crisis of 2008). In other words,
in order to test if our understanding is correct one needs
instances of property bubbles in which there is 
minimal incidence of
exogenous factors. It is from this perspective that the
Hong Kong bubble is of particular interest%
\qfoot{For our research 
another important advantage of the Hong Kong
case was the fact that the website of the ``Rating and Valuation
Department'' of the Hong Kong government provides very
detailed data not only for prices but also for rents.}%
.
\qpar

From its start around 2003 to its climax and burst in the
fall of 2015, the rising phase of the 
Hong Kong residential property bubble offers
so to say an ideal text-book case of a speculative bubble.
Why? In order to answer this question we need to
outline the mechanism through which
property bubbles spring up.

\qA{Mechanism of property bubbles}
The mechanism which leads to a property bubble involves the 
following stages.
\qbu In the initial phase of the process 
low interest rates facilitate purchases by persons who ``need
a roof over their head'' and buy an apartment in order to live in it.
In order to distinguish them from investors and
following a standard terminology
(Roehner 2002, Richmond et al. 2013), these persons
will be called {\it users}.
\qbu As the average price level goes up users are progressively
shut out from the market with the result that the
proportion of the transactions done by investors increases steadily.
Among investors a further distinction is in order between
those who seek a yield (we call them y-investors)
and those who seek a capital gain (we call them c-investors).
\qbu
For c-investors the price level is irrelevant provided they
can resell at a higher price. 
However, because the progression of the rent
is bounded by the income of tenants, the yield 
(i.e. annual rent divided by the purchase price)
is bound to
dwindle as prices shoot up. Thus, y-investors are also
progressively driven out of the market. A market dominated
by c-investors is necessarily
very unstable because any substantial
downward oriented price fluctuation may trigger a 
market downturn.

\qA{Government interventions}

The previous process corresponds to the ideal case where there
are no government interventions. 
The government may try to ``cool off''
the market for instance by increasing interest rates, by
making market access more difficult
for investors who hold already several properties or by increasing
taxes on the profit generated by short-term sales. Moreover,
following the burst of the bubble, the government can prop up
the market by bailing out property developers or by buying
unsold buildings in order to diminish excess inventories.
A good illustration of this kind of anti-cyclic measures is provided
by the policy of the South Korean government from 2007 to 
present (i.e. 2016). In a first phase the measures were aimed
at cooling the market whereas  in a second phase (still
underway) their purpose was to revive it. In recent years
Singapore and China provided two
other examples of government interventions in property markets.
In a general way
government intervention is the rule rather than the exception.
\qpar

On the contrary, in Hong Kong there were few interventions.
As a consequence of Hong Kong's currency board monetary system
(more detailed explanations are given in Appendix A),
the (real) short-term interest rate necessarily follows the US rate. 
Thus, the fact that US rates have been
close to zero over half a decade
provided a fertile ground for the development of the Hong Kong bubble.
As the market downturn occurred only 6 months ago, we do not
yet know if the government (whether the Hong Kong government
or the central government) will come to the help of bankrupt
property developers or whether it will prop up
the market in some other ways. That is why one
should distinguish two
parts in the present paper. 
\qpar

In the first part we analyze the rising phase and show that
almost all the patterns that one expects to observe were indeed
displayed. This includes the following effects.
\qbu The {\it recurrence effect} by which we mean that the 2003-2015
price development was basically the same (in length and rate)
as in the previous episode of 1985-1997.
\qbu The {\it decreasing yield effect} by which we mean that
the rent/price ratio decreased from about 5.6\%  at the
start of the bubble to about 3\% at its peak. Equivalently,
this change
can also be expressed in a way more commonly used in
stock markets by saying that the price earnings ratio (price/rent)
increased from 18 to about 33.
\qbu The {\it price multiplier effect} by which we mean that
for different kinds of apartments
the amplitude of the price peak is an increasing function of
their initial price. The scope of application of the price
multiplier effect is by no means limited to property prices.
It can also be detected in the price peaks of
many other items for which speculation
can take place, e.g. stocks, collectible stamps 
or antiquarian books (Maslov et al. 2003, Roehner 2000,
Roehner 2001 particularly the chapter 
entitled ``Price multiplier effect'').
\qpar

In the second part of the paper
we offer a prediction for the trajectory of the
price fall during the decade 2016-2025, but it must be emphasised
that this prediction rests on the assumption of minimal
exogenous interference.
If there are major changes
in the organization of the market the prediction will no
longer apply.
First, we explain 
our analytical description of speculative peaks. 
This description does not only apply to
property prices but to all
kinds of price bubbles including commodity price bubbles
(see Roehner 2001, p. 158). Secondly, we will see that the values 
of the parameters
$ \alpha $ and $ \tau $ 
describing the rising phases (1987-1997) and (2003-2015) are fairly
similar which leads naturally to the assumption
that they will remain similar for the downward phase. 

\qA{Organization of the Hong Kong property market}
The Hong Kong property market has two components
each of which represents about one half of the market.
\qbu The public housing part is subsidized by the 
Hong Kong government. It offers rental housing
(about 30\% of the whole market) and subsidized sales
(about 18\%).
\qbu Private housing (about 52\%)
\qpar
All the price and rent data used subsequently are for the
private housing part. The coexistence
of a public and private sector is a feature shared by many 
cities, for instance, Singapore or Paris. 

\qA{Changes in the valuation of the housing stock}

What is the global value of residential real estate in Hong Kong?
According to official figures for 2015, there were 3.7 million
persons in private sector apartments, the 
average living space per person of which was 13 square meters
and the average price was about HKD 100,000 per square meter.
(Hong Kong Housing Authority 2015). Thus, one deduces the
following estimate for the private housing stock:
$$ \hbox{\small Private HK housing stock }=
3.7\times 10^6\times 13\times 10^5
=4.8\times 10^{12}= 4.8\ \hbox{\small trillion HKD} $$

As a check one can do an alternative calculation based
on the number of flats. The same report of the Housing Authority
tells us that in 2015 there were 1.5 million flats in the private 
sector. Assuming an average apartment size of 40 square meters%
\qfoot{The average household in the private sector comprises 
only 2.9 persons.}
one gets:
$$ \hbox{\small Private HK housing stock }=
1.5\times 10^6\times 40\times 10^5
=6\times 10^{12}= 6\ \hbox{\small trillion HKD} $$

The public housing sector comprised 1.2 million flats.
However, as their price is not well defined we will limit
our estimate to the private sector. For the sake of
simplicity let us keep 5 trillion HKD as our estimate.
In order to give a clearer interpretation it will be helpful
to translate it into renminbi and USD. With the exchange
rates of May 2016 (1 RMB=1.19 HKD, 1 USD=7.76 HKD) one gets:
$$ \hbox{\normalsize Private HK housing stock (HKhs) }=
4.2\ \hbox{\small trillion RMB }=0.64\ \hbox{\small trillion USD } $$

To get a better sense of the last estimate it can be compared
to the following figures%
\qfoot{The dollar estimate can also be compared with an
estimate of total residential property in
China (including Hong Kong) given
in the ``International Business Times'' (25 January 2016)
which is: USD 39 trillion.}%
:
\qee{1} The revenue of the Hong Kong government 
in fiscal year 2014-2015
was HKD 0.48 trillion (Census and Statistics Department of Hong Kong)
equivalent to ) 0.08 times the value of the
Hong Kong housing stock (at 2015 price level).
\qee{2} The revenue of the Chinese government in 2014
was RMB 6.5 trillion, equivalent to 1.5 times the value 
of the HK housing stock in 2015.
\qee{3} The value of
US Treasury securities held by China at the end of
July 2015  was USD 1.24 trillion equivalent to  1.9 time the 2015
value of the Hong Kong housing stock.

Because the items 1 and 2 are annual flows they must 
be compared with {\it changes}
in the housing stock; for instance one may assume a fall of
10\% per year which would represent 
HKD 0.5 trillion per year which equals
the revenue of the Hong Kong government.
\qpar
In short, a collapse of the HK property market at an annual rate
of 10\%  per year may have a substantial impact not only in Hong Kong
but also in mainland China.
Through its sheer magnitude, such a property crisis 
may also have geopolitical implications. We will come back
to this question in the conclusions.
\qpar

Incidentally, it may be seen that 
the collapse of the previous property bubble was smaller in
magnitude for two reasons: (i) The peak price in 1997 was only
about one half of its value in 2015 (in constant HKD).
(ii) The amplitude of the peak (with respect to its initial
level) was also only about one half its value in the 
second bubble, which means that annual changes were almost 4 times smaller.
 
\qA{Downturn in other Hong Kong real estate markets}

In many countries, for instance the United States,
the cycle of office property is not synchronized with 
the cycle of residential property. However
in Hong Kong the apartment cycle is not only
synchronised with the office cycle, it is also synchronised with both the 
market for retail commercial property and that for
flatted factories%
\qfoot{These are buildings similar to apartments
but built for light industrial activity.
In 2014 only about 200,000 square meters were 
still in use. This area represented 0.4\% of the 
total area of residential real estate. In other words,
for a global assessment this category can be omitted.}%
.
The prices of office real estate started to fall
in the fall of 2015 after having reached a price level 
5.5 higher than the level of 2003 (in constant HKD).
If real estate prices are 
not propped up in some way, the simultaneous downfall
of the residential and office markets may create
great difficulties.

\qI{Recurrence effect}
In previous papers (Richmond 2007, Richmond et al. 2013)
it was shown that:
%
\begin{figure}[htb]
\centerline{\psfig{width=15cm,figure=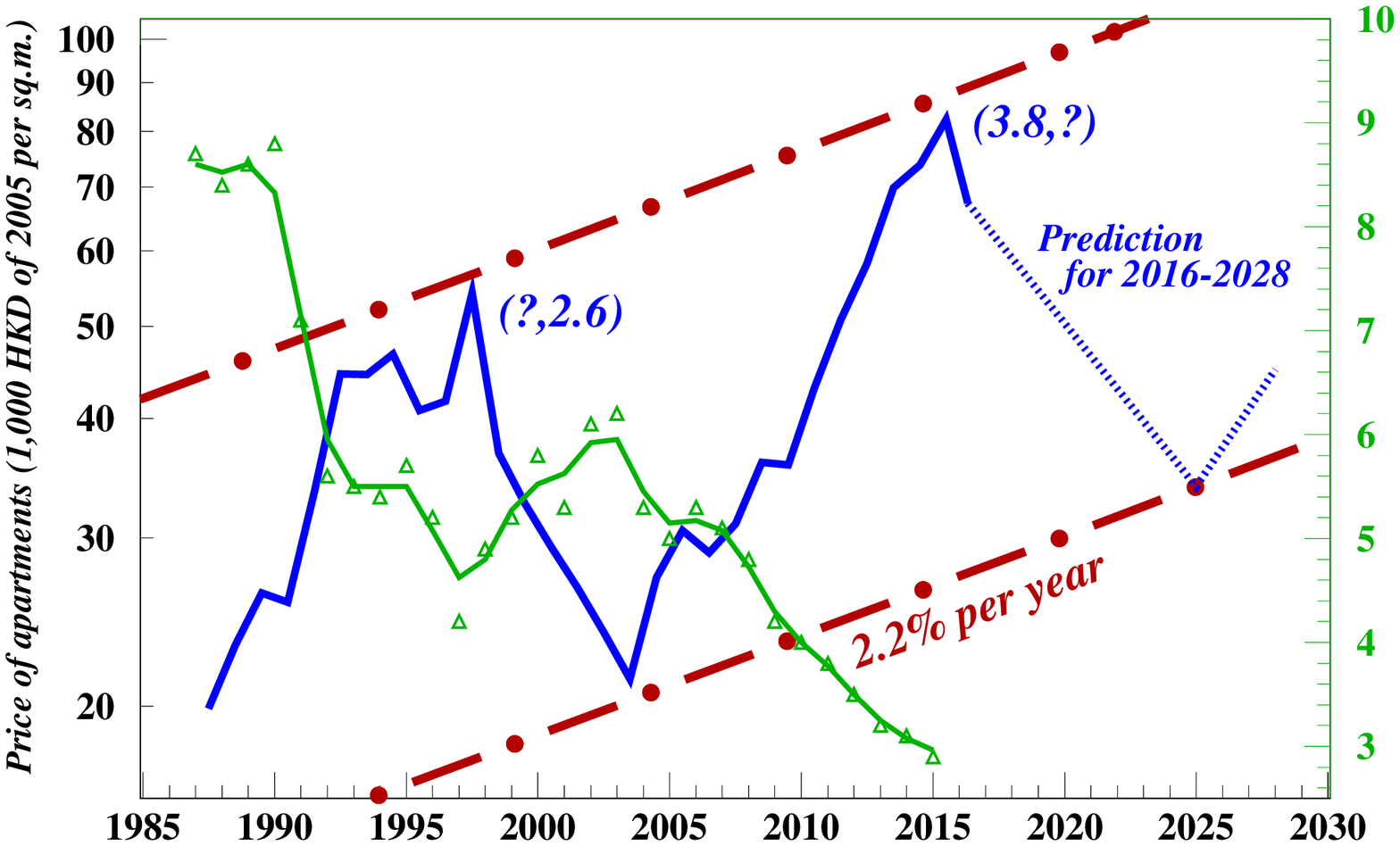}}
\qleg{Fig.\qhu 1\qhv Recurrent speculative episodes in Hong Kong.}
{The prices are annual prices for apartments of less than 40 sq.m in 
the New Territories which constitutes by far the largest part
of the Hong Kong Territory.
The prices of the last quarter of 2015 and
first quarter of 2016 were added in order to show the beginning
of the downturn.
This size-class was selected because
it is likely to have the largest number of transactions
and more transactions will result in a smoother curve; although
somewhat different in price levels, the curves for
the other classes are parallel to this one.
It can be noted that between 2003 and 2015 the wages were 
practically stagnant. 
The dotted curve gives a rough projection based on the first
episode. A more precise prediction will be given below.
The numbers within parenthesis give the amplitudes of the
upward and downward phases of the respective peaks.
The green descending line shows the annuals yields (scale
on the right-hand side expressed in percent); 
they have a negative correlation
of -0.76 (the 0.95 confidence interval is -0.88,-0.54) 
with the prices.}
{Source: Rating and Valuation Department of the
Hong Kong Government. ``Trading Economics'' website
for the evolution of wages.}
\end{figure}
%
\qbu Between 1970 and 2015 there were three recurrent property bubbles
in the UK and two in Ireland.
\qbu Between 1960 and 2015 there were four recurrent property bubbles
in the west of the United States.
\qL
The corresponding price peaks followed one another in such a way
that the end of the downward phase of episode number $ k $ was immediately
followed by the start of the upward phase of episode number $ k+1 $.
Moreover the shapes of the prices peaks were basically the same.
\qpar

Similarly, Fig. 1 shows that between 1983 and 2016 
Hong Kong experienced two speculative episodes in close succession.
It shows also that the current bubble is of greater magnitude
than the previous one.

\qI{Decline of the yield, increase of the PER}

From the perspective of an investor
whether or not an apartment or a stock is overvalued
is determined by the yield. For stocks it is the 
ratio of the annual dividends to the price of the stock.
For a house or an apartment it is the ratio
of the annual rent to the price. The price earnings ratio
(PER) is the inverse of the yield.

%
\begin{figure}[htb]
\centerline{\psfig{width=17cm,figure=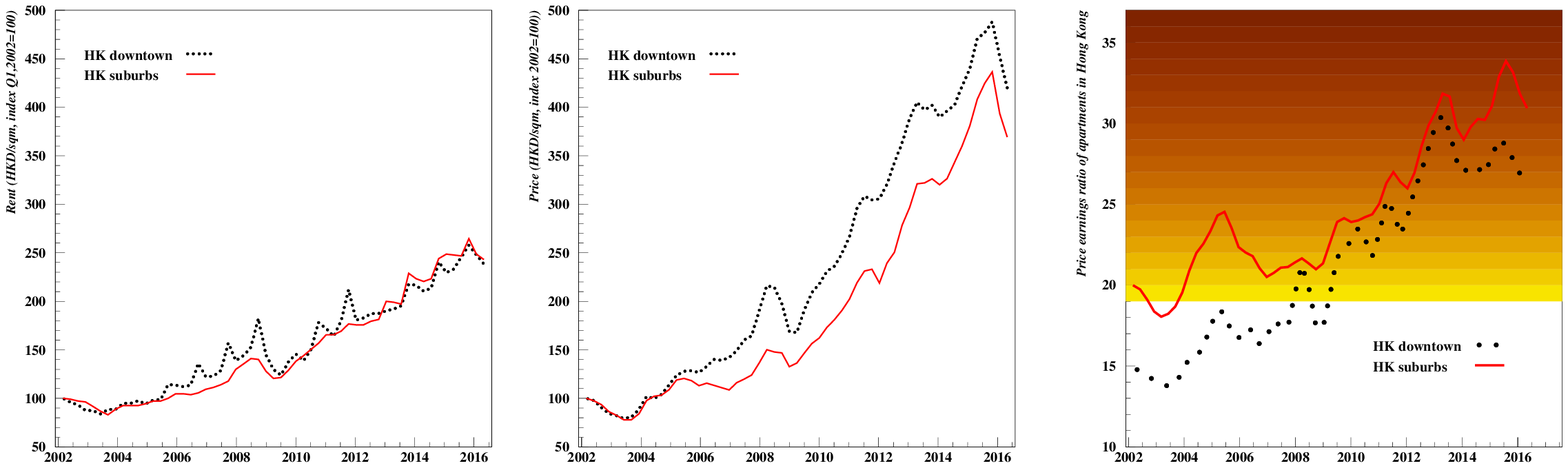}}
\qleg{Fig.\qhu 2\qhv From left to right (i) rent index, 
(ii) price index and (iii) price earnings ratio.}
{``Downtown'' means the Hong Kong island; ``suburbs'' means
the ``New Territories''. The data are for apartment of
less than 40 square meters and are expressed in current HKD.
A PER of 20 means a yield of 1/20=5\%.}
{Source: Rating and Valuation Department of the
Hong Kong Government.}
\end{figure}

For stock markets
historical records show that the long-term average
of the PER is comprised between 15 and 20. A PER of 35
indicates a fairly unstable market which is bound to fall
sooner or later. Historical data for property
markets lead to the same conclusion.
The PER reached a level
of 35 in the ``New Territories'' in the fall
of 2015, thus signaling a notable market instability.
\qpar

A PER of 35 corresponds to a yield of 2.8\%. Note this is a
a gross rental yield in the sense that it does not make any
allowance for vacant periods, administration costs,
repairs, property taxes. As a result, 
it is safe to say that landlords in Hong Kong earn essentially nothing on their
apartments or may even have to accept a negative yield. 
\qpar

Fig. 2 shows that prices increased faster than rents particularly
in the second half of the ascending phase.
It also shows that for the more expensive downtown apartments
the relative price increase was higher. This constitutes
the ``price multiplier effect'' that will be considered in
more detail below. On the contrary, the increase in rent
was the same in downtown and suburbs.

\qI{Fall in sales versus fall in prices}

It is commonly observed that the volume of transactions
begins to fall before prices start to fall.

%
\begin{figure}[htb]
\centerline{\psfig{width=9cm,figure=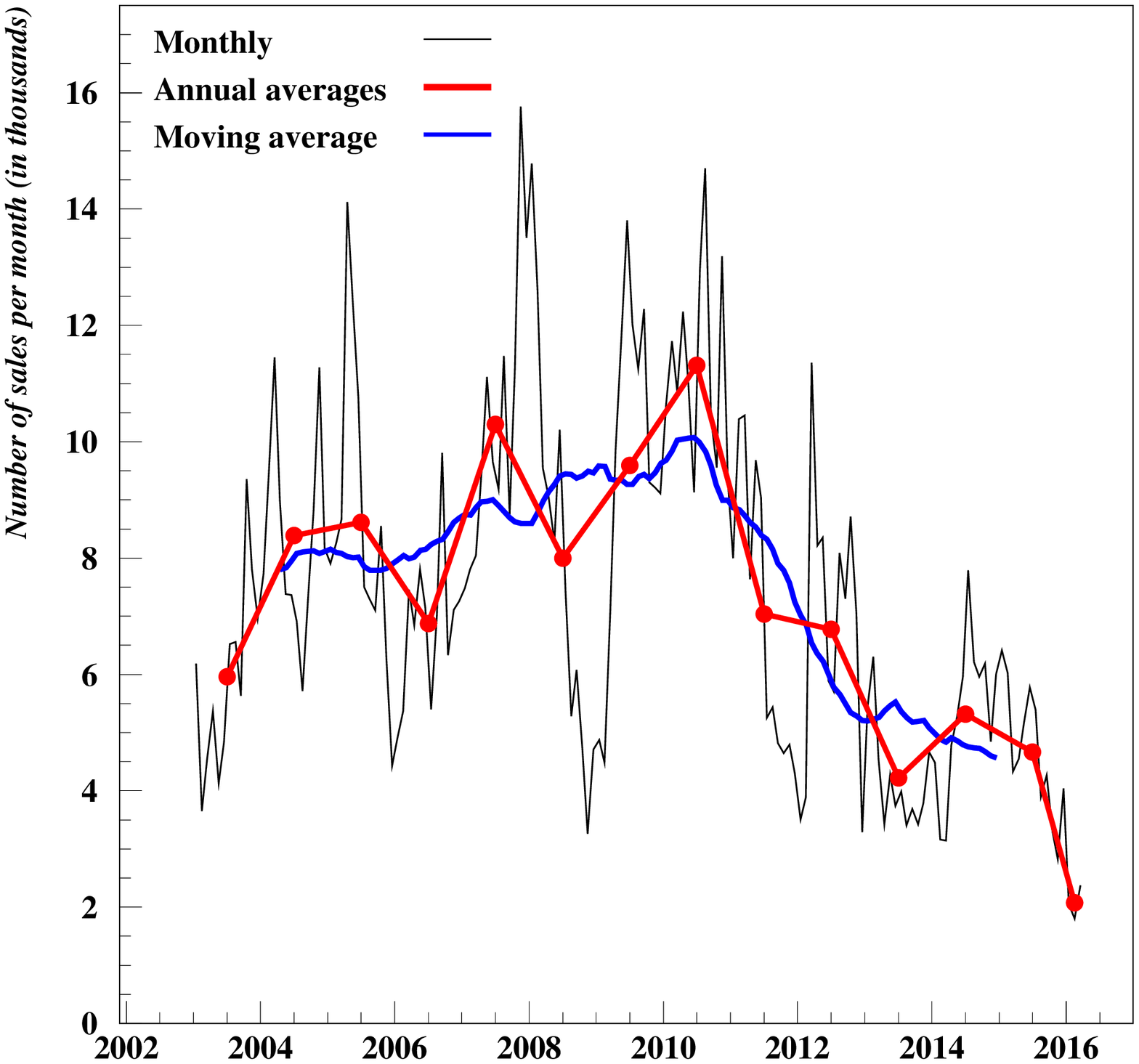}}
\qleg{Fig.\qhu 3\qhv Monthly number of sales.}
{The curve with the red dots shows the annual
averages of monthly sales. The blue curve without dots
corresponds to a moving average window of 31 months.}
{Source: Rating and Valuation Department of the
Hong Kong Government.}
\end{figure}
As can be seen in Fig. 3 monthly sales are highly volatile.
For the picture to become more meaningful one must apply
some kind of averaging. Although less volatile,  annual averages
do not indicate any clear trend until the fall becomes
really substantial. By using a moving window with a width of
31 months one sees a clearer trend albeit with the drawback that
at each time point the last 15 months will be hidden.
It means that the trend will be displayed with a 
time lag of 15 months. Thus, for the purpose of forecasting
the downturn of prices, the moving average technique is hardly
better than the annual sales.
\qpar
Sales began to fall in late 2010 that is to say
5 years before prices also eventually began to fall. Such a long
gap is fairly uncommon and may probably be due to the
very low interest rate over the period 2010-2016.

\qI{Test of the price multiplier effect}

During a speculative episode the price of an item jumps
from an initial level $ p_1 $ to a peak level $ p_2 $ before more or less
returning to level $ p_1 $. 
The ratio $ p_2/p_1 $ is referred to as the amplitude $ A_1 $ of the
peak. 
It turns out that usually $ A_1 $  is an increasing function of the
logarithm of the initial price (Roehner 2001, chapter 6):
$$  A_1=m\ln p_1 +b   \qn{1} $$

There are two ways in which 
we can test whether this regularity holds or not. 
One is with respect to location, i.e.
Hong Kong versus New Territories, the other is with respect
to apartment size.

\qA{Price and amplitude with respect to location.}
In the first quarter of 2003, the price per square meter
of apartments of less than 40 sq.m. was kHKD 21.3 in the
new Territories and 25.3 in Hong Kong island.
There was a similar price gap in the other size classes.
On average over all size classes, the
New Territories had a price of kHKD 27.0 against 41.5 for
Hong Kong island. Therefore, according to the price
multiplier rule, one would expect a higher amplitude
for the Hong Kong prices than for the New Territories prices.
And indeed we see below that this is the case.
\qpar

For the 5 size classes, the parameter $ m $
in relationship (1) is positive with the following values:
$$ (0,40):\ 3.9\quad (40,70):\ 1.4\quad (70,110):\ 1.5\quad 
(110,160):\ 1.9\quad (>160):\ 3.0 $$

The average over all size classes is $ \overline m=2.3 $
which is consistent with the values found for different
locations in other markets (Roehner 2001).

\qA{Price and amplitude with respect to size}
In Hong Kong island, in the first quarter of 2003, 
the price per square meter
of apartments of more than 160 sq.m. was kHKD 65,
that is to say 2.5 times more than the price of 25.3 recorded
in the smallest size class.  If one takes into account
the other size classes, one finds a high correlation of 0.99
between size and price per square meter.
Therefore, according to the price
multiplier rule, one would expect a higher amplitude
for the large apartments than for the small ones.
\qpar

But now find surprisingly that it is exactly the opposite.
There is indeed a significant correlation between 
the amplitude and $ \log(p_1) $ but instead of being
positive as expected, $ m $ is actually negative:
$ m=-1.8\pm 1\quad  (\hbox{correlation of } -0.89) $.
\qpar

The same observations hold for the New Territories.
There is again a high correlation of 0.98 between
size and price but $ m $ is also negative:
$ m=-5.2\pm 1.4 $ (correla\-tion of  $ -0.97 $) 
\qpar

Expressed in words, the larger the apartment, the less
active was speculation. One possible reason for this couple be the existence of fiscal regulations aimed at discouraging luxury purchases.

\qI{Prediction for the price trajectory 2016-2025}

In Fig. 1 we gave a rough prediction which was simply based
on the symmetry argument. In this section we propose
a more precise description of the shape of the downward price
trajectory. In line with previous publications
(Roehner 2001 chapter 7, Richmond et al. 2007) 
a price peak will be
described by the following two-parameter function
($ \alpha $ and $ \tau $ are the two parameters 
whereas $ t_2,p_2 $ are given by observation):
$$ p(t)=p_2\exp\left[-\left|{t-t_2 \over \tau}\right|^{\alpha}\right] 
\qn{2} $$

The top of the peak is defined by time $ t_2 $ and price $ p_2 $.

%
\begin{figure}[htb]
\centerline{\psfig{width=17cm,figure=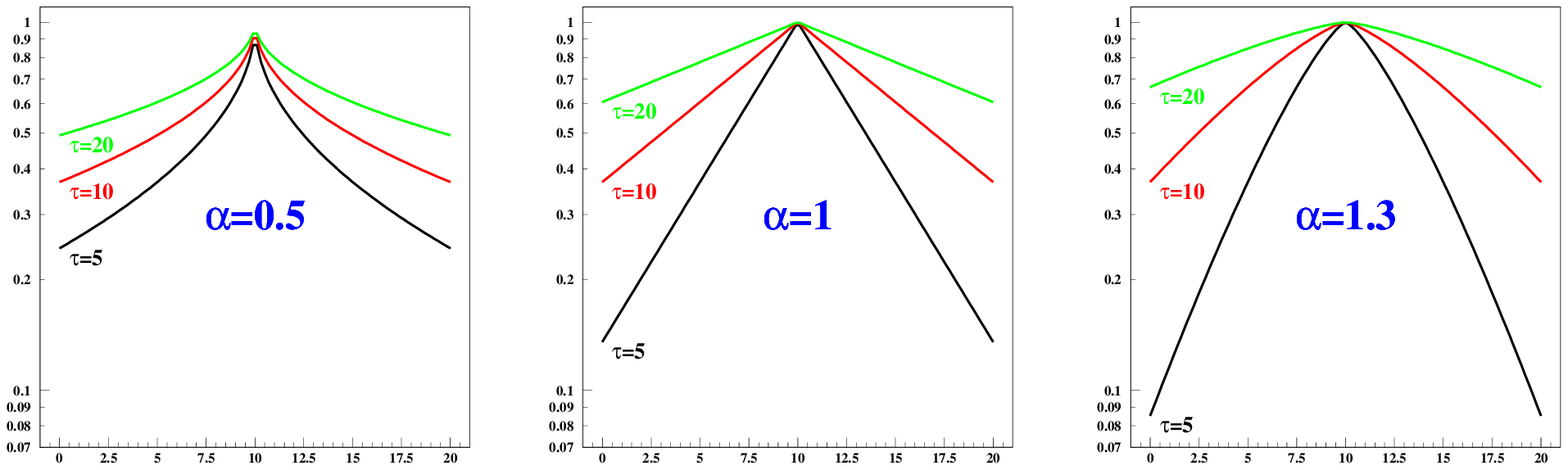}}
\qleg{Fig.\qhu 4a\qhv Role of the parameters $ \alpha $ and
$ \tau $.}
{The parameter $ \alpha $ describes the sharpness of the 
peak near its summit whereas the parameter $ \tau $ 
defines the overall broadness of the peak.}
{}
\end{figure}

$ \alpha $ determines the shape of the peak
and $ \tau $ determines how fast the price
decreases on each side of the peak.
\qpar

The estimation procedure of the parameters consists in two steps.
\qbu Linearization of the rising and declining sections \qbu Least square regressions which gives $ (\alpha_1, \tau_1) $ 
and $ (\alpha_2, \tau_2) $.
\qpar

Then, the prediction procedure involves three stages.
\qee{1} First we estimate the parameters for the peak 
1987-2002 using quarterly prices; this leads to
(the $ \tau $ are expressed in years): \qL
upward phase: $ \alpha_1= 0.80,\quad \tau_1=12 $ \qL
downward phase: $ \alpha_2=0.41,\quad  \tau_2=16 $\qL
The fact that $ \alpha_2<\alpha_1 $ shows that
in the vicinity of the summit
the declining trajectory is  steeper than the rising
trajectory.
\qee{2} Secondly, we estimate the parameters of the
upward phase of the second peak (also using quarterly
prices) in order to check if they are similar to those
of the first peak; this leads to:
$ \alpha_1=1.1 , \quad \tau_1=10 $\qL
For $ \alpha_1 $ the difference with respect to the previous 
pair $ (\alpha_1,\tau_1) $
is about 30\% and 18\% for $ \tau_1 $. 
%
\begin{figure}[htb]
\centerline{\psfig{width=9cm,figure=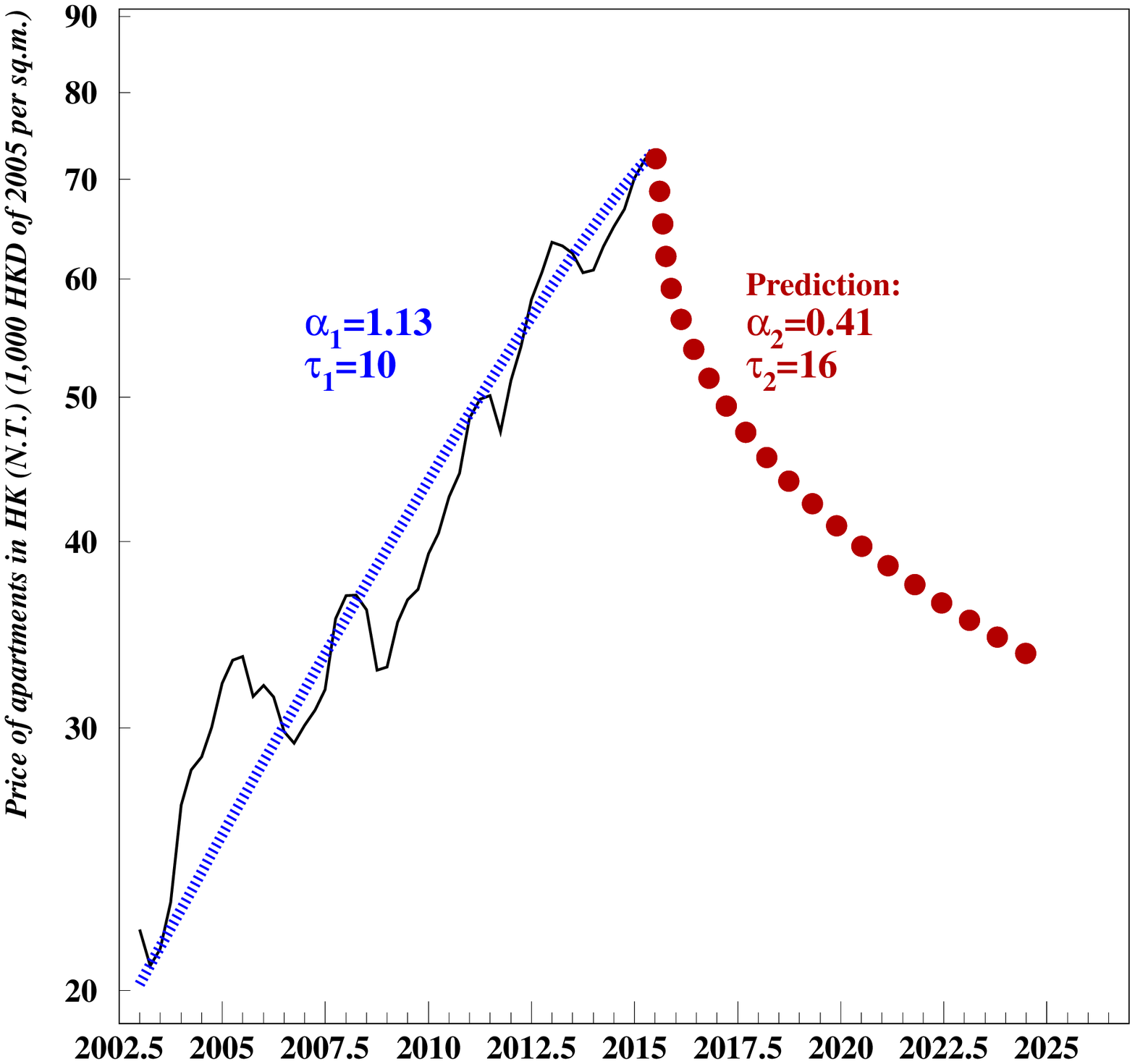}}
\qleg{Fig.\qhu 4b\qhv Predicted downward trajectory
2016-2025.}
{The prediction is based on the assumption that
this speculative episode will be similar to the
episode of 1987-2002. This assumption is supported
by the fact that the upgoing phases are fairly similar.}
{}
\end{figure}
%
\qee{3} As the previous step gave confidence in a
similarity of the two episodes, the predicted
downward trajectory was drawn by using the 
same parameters as in the downgoing phase of the 
first episode.
This led to Fig. 4b.

\qA{Property bubble in London?}

A report by the Swiss bank UBS (Holzhey and Skoczek 2015)
issued in October 2015 gave Hong Kong and London as the 
two most overvalued real estate markets worldwide.
It is therefore natural to briefly compare the two
markets.
%
\begin{figure}[htb]
\centerline{\psfig{width=10cm,figure=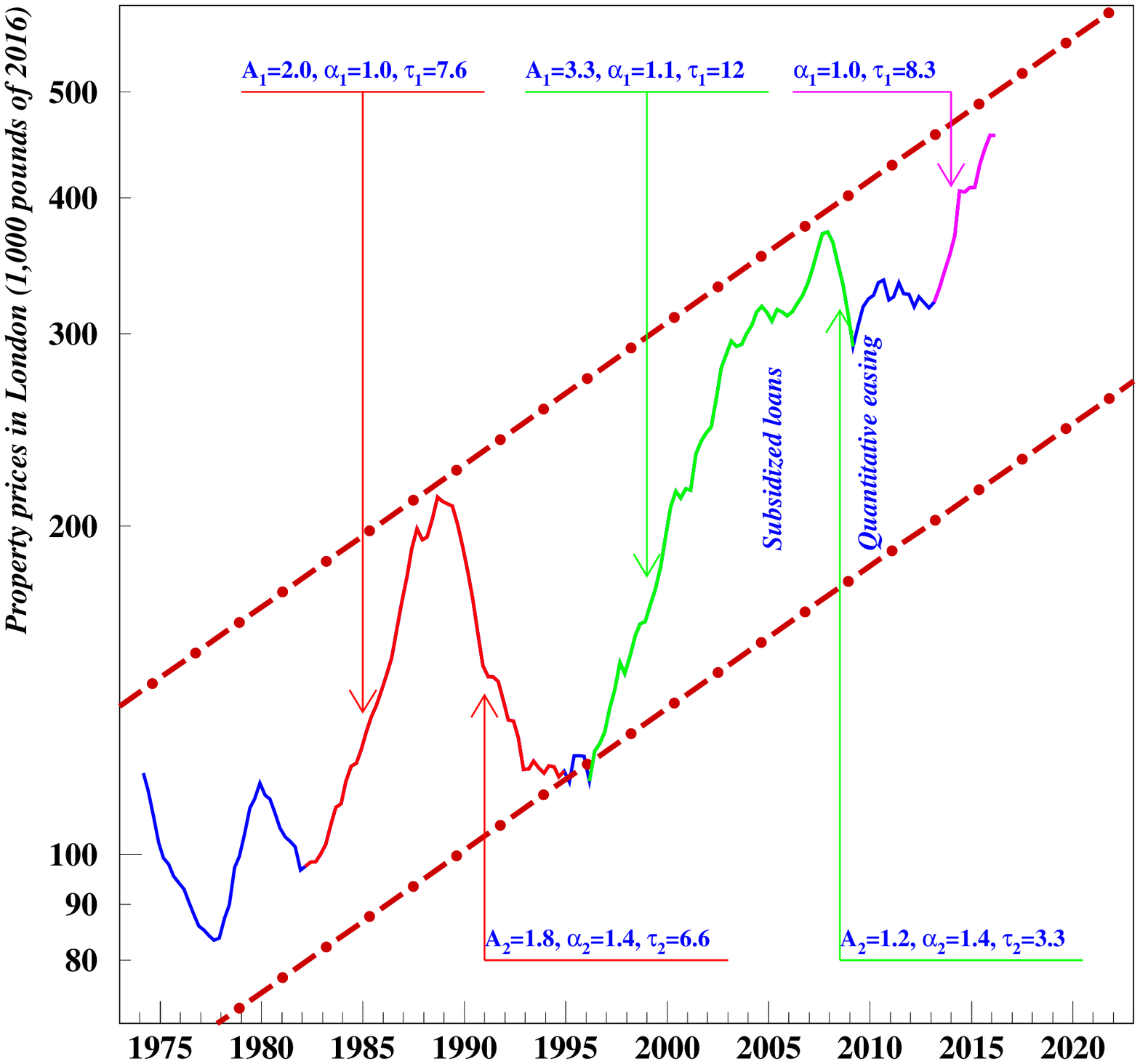}}
\qleg{Fig.\qhu 5\qhv Price of apartments in London.}
{The recurrent nature of these speculative episodes
is emphasized by the stability of the shape parameter
$ \alpha $: it is around $ 1 $ in the
ascending phases and around $ 1.5 $ in the
descending phases.
According to the ``Nationwide'' press release of July 2016, 
the Brexit vote of 23 June 2016 hardly affected
UK house prices: the change from June to July
was the same as from May to June, namely an 
(annualized) increase of 5.2\%. Whether the same
stability will prevail in coming months
remains an open question.}
{Source: The data are from the website of the
``Nationwide Building Society''.
}
\end{figure}

Fig. 5 suggests that the situations
in Hong Kong and London may be very different
notwithstanding the UBS ranking.
More precisely, in 2008 London was in the same situation
as Hong Kong in 2015 in the sense that 2008 marked
the end of a long period of price increase which
had brought about a multiplication of prices by $ A_1=3.3 $
between 1996 and 2008.
As a matter of fact, this rising phase was characterized
by an $ \alpha $ value which was very similar to the 
$ \alpha $ of the upward phase of 1983-1990.
Moreover in terms of shape,
the declining phase which started in 2008 was also
similar to the previous one of 1990-1996, with the same
$ \alpha $ value of $ 1.4 $ in both cases.
\qpar

In 2010, however, everything changed. The trend reversed
and a new upward phase started. Whether or not this was due
to the policy of ``quantitative easing'' is difficult to say.
It is clear that this policy made it easier for
banks to provide loans but by itself this should not 
create an appetite for real estate operations if the
market is perceived as overvalued. The UBS report
emphasizes that with
a rent/price yield under 3\% the market was indeed somewhat
overvalued at the end of 2015.
\qpar

The Brexit vote of 23 June 2016 
added an additional  
layer of uncertainty. Not surprisingly, the ``Nationwide'' press
release of July 2016 says that
 ``the housing market outlook is unusually
uncertain''. As a matter of fact, the massive exogenous factors
of quantitative easing (or the end of it) and Brexit
make any prediction impossible at least in the framework of our
model. If we assume that the rising phase of 2014-2016
was the continuation of 1996-2008, then one is in a 
Hong Kong like situation. On the contrary, if one assumes that
a new cycle started in 2010, then the market appears in
a completely different perspective.

\qI{Conclusion}

It has been emphasized that in its ascending phase which covers 2003-2015
the Hong Kong residential property market followed the general rules
and characteristics of speculative episodes in property markets
that came to light through the analysis of episodes
which occurred in other countries.
One can keep in mind the following features:
\qbu The recurrence rule means that the characteristics of the
present episode are fairly similar to the ascending phase (1986--1997)
of the previous Hong Kong epi\-so\-de.
\qbu The yield rule (or equivalently price earnings ratio rule)
described the erosion of the yield which fell from
6.2\% in 2003 to 2.9\% in 2015.
\qbu The price multiplier effect underlines the fact that the
amplitude of the price peak is highest for the most
expensive segment of the property market.
\qpar
Based on the previous regularities we proposed a testable prediction
of the price trajectory for the interval 2015-2025.
\qpar

The predictions proposed in Roehner (2006) and in Richmond (2007)
respectively
for the west of the US and for London-Dublin 
turned out to be reasonably successful%
\qfoot{A comparison between the predictions and 
the actual price trajectories was made in Richmond et al. (2013).}%
.
In 10 years from now on it will be interesting to see
how successful the present prediction is.

\qA{What will be the consequences for the Hong Kong economy?}

Will the property slump lead to bank failures? To what
extent will it affect the economic growth of Hong Kong?
Will the crisis spread to mainland China? 
The only way to give ``experimental''
answers is to analyze what happened in the previous episode. 
\qpar

In answer to the first question it turns out (rather surprisingly)
that there were no major bank failures as a result of the
property slump. According to the Monetary Authority
in August 2003 (that is to say at the end of the slump)
only ``22\% of all residential mortgages were larger
than the current value of the properties they financed''
(Bradsher 2003).
How is such a low percentage compatible with the fact
that property prices had been divided by 2.6 between 1997 and
2003? The only possible explanation seems to be that the
loans covered only a fraction of the price.
In the same article one learns that banks required borrowers
to put up at least 30\% of the price of the apartment. 
Bradsher suggests that the persons
who bought an apartment in fact borrowed less than 50\% of 
its price. That is why, it is the 
Hong Kong middle class, not the banks, which bore the
brunt of the crisis. In this respect, one should 
recall that in this period all Asian countries had a 
very high saving rate. Will the present bubble again be
absorbed by the buyers rather than by the banks? 
\qpar
There are several indications to the opposite.
\qee{1} In order to attract more middle class customers
the banks offered loans over very long durations
(up to 30 years) and which covered up to 90\% or
even 100\% of
the price.  Most of the loans are variable rate loans
which means they can be adjusted for inflation.
Thus, one would suspect that once price will have fallen
substantially many people will become unable to
repay their loans, not so much the interest but rather
the capital itself.
\qee{2} A confirmation of the  previous characteristics
can be found in present-day Internet 
advertisements of Hong Kong banks. For instance HSBC 
( ``Hong Kong and Shanghai Banking Corporation'') and the 
``Development Bank of Singapore''
(DBS) propose 
mortgages with a loan-to-value ratio of up to 90\%.
Every bank offers its own mortgage 
rates and the HK government
does not publish any average. Comparison of loan conditions
is made even more difficult because, as in the US prior to
the subprime crisis, the banks offer ``perks'',
for instance a free fire insurance.
\qpar

At a more macroeconomic level, Table 1 compares the performances
of Hong Kong, China, Singapore and South Korea during the
1992-2002 time interval.

\begin{table}[htb]

\small

\centerline{\bf Table 1\quad Comparative growth in
Hong Kong and neighboring countries before and after 1997}

\vskip 5mm
\hrule
\vskip 0.7mm
\hrule
\vskip 2mm

$$ \matrix{
 & \hbox{Real GDP} & \hbox{Real GDP} & \hbox{Ratio} & 
\hbox{CPI} & \hbox{CPI} \cr
 & 1992-1997 & 1997-2002 &  & 1992-1997&  1997-2002\cr
\qtb
 & g_1 & g_2 &  g_2/g_1 & & \cr
\noalign{\hrule}
\qth
\hbox{Hong Kong} \hfill & 12\% & 6.5\% & 0.56 & 9.2\% & -2.9\% \cr
\hbox{China} \hfill & 12\% & 8.0\% & 0.67 & 13\% & 0.64\% \cr
\hbox{Singapore} \hfill & 12\% & 3.4\% & 0.28 & 2.5\% & 0.55\% \cr
\qtb
\hbox{South Korea} \hfill & 9.3\% & 5.7\% & 0.61 & 5.0\% & 3.9\% \cr
\noalign{\hrule}
} $$
\vskip 1.5mm
\small
Notes: GDP means ``Gross domestic Product'', real GDP means
corrected for inflation; 
CPI means ``Consumer Price Index''; the figures
are average annual changes.
In Hong Kong and Singapore there were
property bubbles which burst in 1997; the collapse of real
estate prices was
faster in Singapore than in Hong Kong.
\qL
{\it Source: Website of ``Trading Economics''.}
\vskip 5mm
\hrule
\vskip 0.7mm
\hrule
\end{table}

Three features appear fairly clearly.
\qbu Growth was everywhere faster before 1997 than after. 
This was 
due partly to the impact of the crisis of 1997-1998;
growth was slowest in Hong Kong and Singapore which 
in addition experienced real estate crashes.
\qbu There was a marked deflation in Hong Kong; in addition
real wages (not shown) remained stagnant.
\qbu In Singapore real estate prices fell by 50\% within two years,
that is to say twice as fast as in Hong Kong.
This sharp fall was probably the main cause of the sluggish growth.

\qA{Property bubble in Taipei}

In 2016 the Taipei-Keelung-Taoyuan metropolitan area in the 
north of Taiwan had a population of 9.1 million
which represented 40\% of the
total population of Taiwan. Since 2001 it has developed a speculative
bubble which, for the whole of Taiwan, has an amplitude of 2.5
once adjusted for inflation. Although 40\% smaller
than in Hong Kong%
\qfoot{Perhaps due to the fact that from 2011 to 2015 the interest
rate had remained near 2\% compared to almost zero in Hong
Kong.}%
,
it may nevertheless result in a serious property slump in coming years.
The summit of the price peak was reached in May 2015. 
The parameters of the upgoing 
phase were $ \alpha_1=1.1,\ \tau_1=11 \hbox{year} $, 
not much different from the values in Hong Kong, namely
$ \alpha_1=1.1,\ \tau_1=13 \hbox{year} $.

\qA{Possible geopolitical implications of a property slump in Hong Kong}

Over the past ten years real estate prices did increase markedly
not only in Hong Kong but also in the centers of major
Chinese cities such as Beijing, Guangzhou or Shanghai.
However there were two main differences: (i) In the mainland
wages and salaries were multiplied by two or three whereas in 
Hong Kong they stagnated. (ii) If one includes 
the suburbs of major cities, the inland
cities (e.g. Chongqing) or smaller cities
(not to speak of rural areas), then the average price increase 
is much smaller than in Hong Kong. That is why the burst of the
bubble is expected to have more detrimental effects in 
Hong Kong than in the mainland. 
\qpar
In its February 2016 assessment of the situation in
Hong Kong the Fitch rating agency sees the greatest risk
for Hong Kong banks in their exposure to mainland loans.
Bearing in mind the 
magnitude of the property assets in Hong Kong (of the order 
of one or two trillions of USD) we propose turning this 
proposition around and asking the following questions:
\qbu What is the exposure of mainland banks to Hong Kong
property loans?
\qbu Will the Beijing government have the will and capacity
to bail out Hong Kong?
\qbu If Hong Kong is indeed bailed out by the Chinese government
will that change the relationship between Hong Kong and the
Chinese Government.
\qpar

Giving a reliable answer to the first question would
require a specific study that would be out of place in the
present paper. Trying to answer the second question would
be a fairly speculative matter. Therefore we will focus
our attention on the third question. More precisely, will a bailout
improve or worsen
the relations between Beijing and Hong Kong?
\qpar

At first sight it might seem that a bailout should earn
Beijing some gratitude, however the recent example
of Greece versus Germany shows that if the bailout is done
with contempt and arrogance it may worsen the relation
between the partners.
\qpar

Back in 1997 the so-called pan-democracy camp
adopted a fairly non-cooperative stance with respect to the Chinese 
government. The fact that overall Beijing had respected
the ``Basic Law'' did not close the gap. On the 
contrary, during the past decade several movements
and parties have appeared whose program contains
claims for independence%
\qfoot{One can mention the following (the date of creation is
given within parentheses):
\qbu The ``Hong Kong Independence Movement'' (Feb. 2005)
\qbu The ``Hong Kong Autonomy Movement'' (May 2011)
\qbu ``Civic Passion'' (Feb 2012)
\qbu The ``Hong Kong Independence Party'' (Apr 2015)
\qbu The ``Hong Kong National Party'' (Mar 2016).
\qpar
Obviously,
there has been an acceleration in the creation of such groups
in recent years. It can be observed that the very existence
of secessionist movements is in contradiction with Article
23 of the ``Basic Law'' which plays the role of the constitution
of Hong Kong.
Paradoxically, in their demand for more sovereignty, these groups
do not seem to mind the fact that the Hong Kong interest rate
is bound to follow the US rate.}%
.
\qpar
Here again it would be mere speculation to try to
predict what will be the attitude of the central
government for we do not have any precedent which could
serve as a guide%
\qfoot{However there was a small-scale precedent in 1985
when the shipping business of Tung Chee-hwa
was rescued from looming bankruptcy by a USD 100 million
loan by the PRC. In 1997 Tung became the first
Chief Executive of Hong Kong.}%
.
However, it is worthwhile to keep in mind
possible implications of the crisis in Hong Kong.
Will the Chinese government be able to prevent a domino effect?
Will it be clever enough to take advantage of this opportunity
to tighten the links with Hong Kong? Moreover what will
be the policy of the United States? In broad lines it was
defined in the ``Hong Kong Policy Act''  passed in 1994%
\qfoot{This Act ties the benefit of specific economic
privileges granted by the US to HK
to the preservation of a large autonomy.}%
which continues to raise Chinese displeasure because
by favoring Hong Kong's autonomy it
set up a soft form of interference.

\appendix

\qI{Appendix A: Role of the USD--HKD currency board}

The HKD is pegged to the USD in a currency board system.
The verb ``to peg'' means to fix at a certain level. The peg
between the US dollar and the Hong Kong dollar means that 
the exchange rate between the two currencies is to remain
fixed at a given level, currently 7.80 HKD for one USD.
\qpar

Currency pegs are fairly common.
At one time there was
also a peg between the renminbi and the dollar.
However the Hong Kong peg is special in two respects.
\qbu In  contrast to the Singapore peg which is with 
respect to a 
(confidential) basket, of various currencies, HK has
a fairly rigid peg with only one currency.
\qbu A currency board system is quite uncommon
because under such a system
HK's major monetary decisions (for instance the
volume of money supply) are US-dependent.
\qpar

In financial publications the monetary
system of HK
is usually praised for having ensured a rapid
development and a great stability. However, when one
compares Hong Kong to Singapore this judgment appears to
be somewhat overrated: between 1976 and 2010, the real
GDP of Singapore was multiplied by 10 against 5.6
for Hong Kong; in terms of stability Singapore had also
a better achievement. For instance from 1989 to 1996, 
Hong Kong had an inflation
rate over 10\% whereas in Singapore it remained below
5\%.

%
\begin{figure}[htb]
\centerline{\psfig{width=17cm,figure=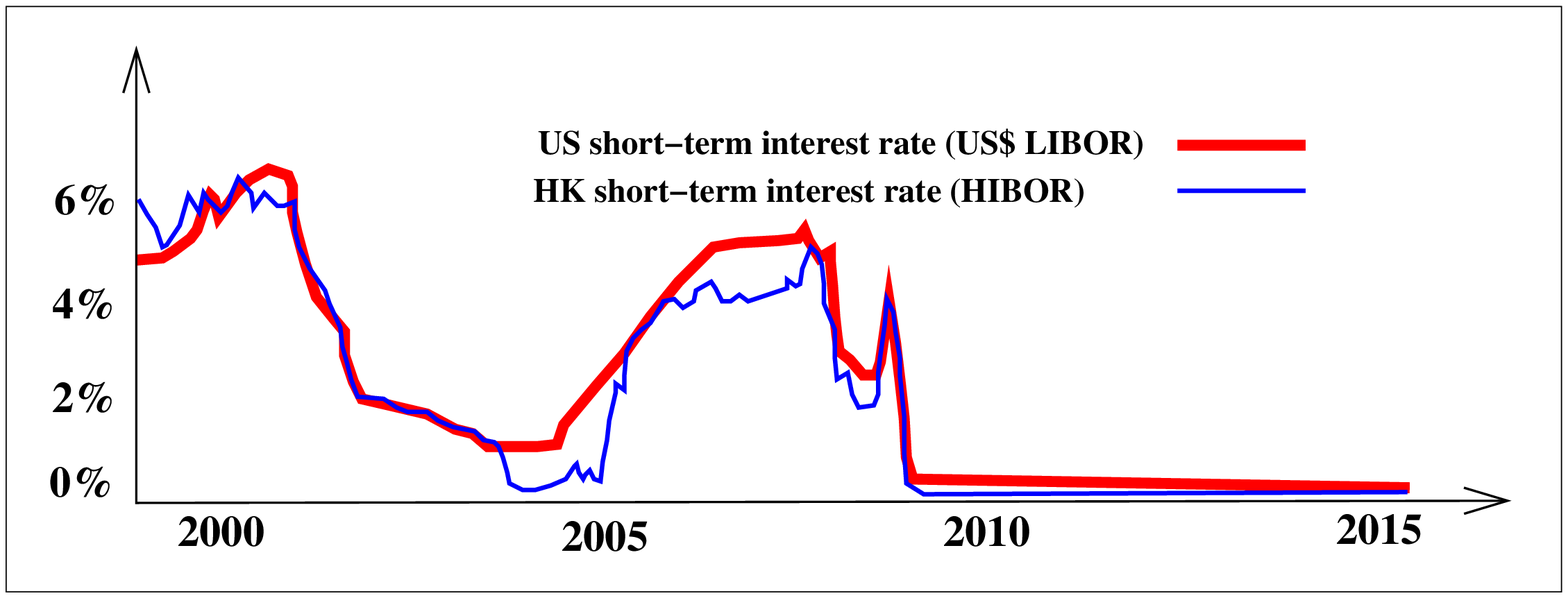}}
\qleg{Fig.\qhu A1\qhv Close link between US and Hong Kong
short-term interest rates.}
{LIBOR (which means London Interbank Offered Rate) consists
of a whole set of series respectively for the dollar, British 
pound, euro, and so on. HIBOR (which means
Hong Kong Interbank Offered Rate) is a short-term 
rate for the HKD. This high correlation is an effect of the
HKD-USD peg. After the initial decision in 1983 the
connection was made tighter in subsequent years due to
institutional improvements such as the ``Accounting arrangements''
of 1988: from 0.70 in the time interval 1984-1988 the
correlation between interest rates rose to 0.98 in 1988-1994.}
{Sources: Pershing Square 2011, Trading Economics: Fed funds rate
versus HK benchmark rate, Hong Kong Monetary Authority 1994.}
\end{figure}

%
\begin{figure}[htb]
\centerline{\psfig{width=9cm,figure=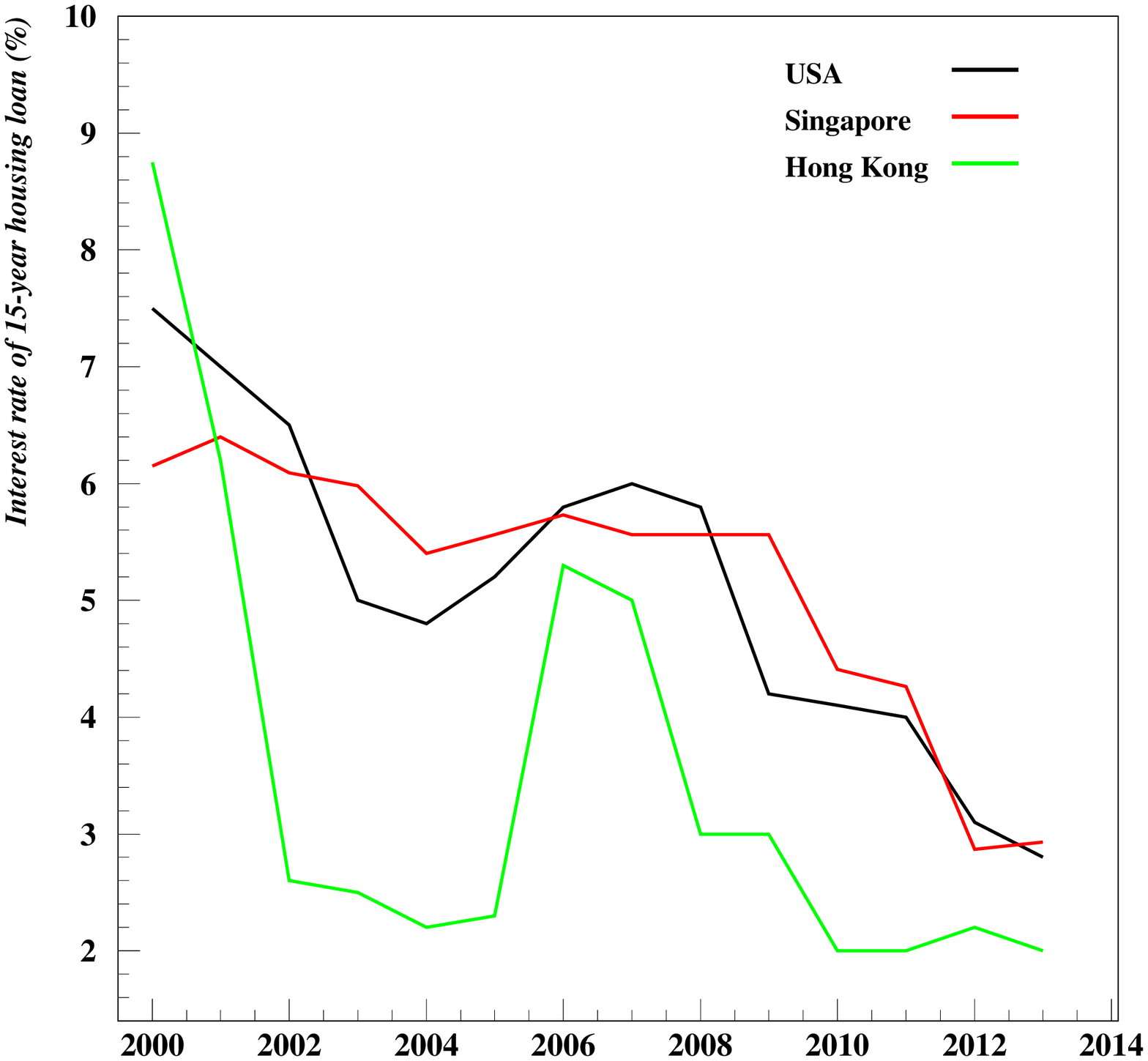}}
\qleg{Fig.\qhu A2\qhv Mortage rates in the USA, Singapore and
Hong Kong.}
{Unsurprisingly,  with respect to short-term rates,
mortgage rates have their own dynamic.
The correlation USA--Singapore is 0.87 whereas the
correlation USA--Hong Kong is 0.77.
Between 2009 and 2016 the inflation rate in Hong Kong
was around 4\%. As the average mortgage rate was lower
than inflation, one wonders how selling mortgages
could be profitable. In fact, one should keep in
mind that every HK bank offered its own specific loan
conditions.}
{Sources: 
USA: https://ycharts.com. 
Singapore: Monetary Authority of Singapore.
Hong Kong: Based on figures for new mortgages from the 
Hong Kong Monetary Authority's monthly ``Residential
Mortgage Survey''.}
\end{figure}

The important point is not the peg itself for indeed
many currencies are pegged in some way or another.
For instance, since 1985 the Singapore dollar is
pegged to the central parity of an
undisclosed trade-weighted basket of currencies 
and
is allowed to float within an undisclosed bandwidth 
of this central parity. Actually a peg is never completely
rigid. Thus the HKD is allowed to float within the interval
$ 7.80\pm 0.05 $. It is only when one of these limits
is reached that the Monetary Authority (Hong Kong's {\it de
facto} central bank) intervenes.
\qpar

We said that in addition to the peg, Hong Kong 
has also a currency board. The implication 
of such a monetary system is that 
in order to issue new HK banknotes the note-issuing banks%
\qfoot{Two of the  three note-issuing banks are British, namely
HSBC and
the ``Standard Chartered Bank'' and one is Chinese (Bank of China).}
must submit the same amount 
in USD (at the fixed exchange rate) to the ``Exchange Fund''.
Through this process the Exchange Fund has accumulated
a huge amount of dollars; at the end of 2015 it had reached
\$440 billion. 
\qpar

Fig. A2 shows that with respect to short-term rates
mortgage rates have their own
dynamic. However, very low short-term interest rates
made it more difficult for the 
HK Monetary
Authority to ``cool the market'' by raising mortgage
rates substantially (assuming it was sufficiently
independent from the wishes of the banking sector to 
have the will to do that, which is not obvious).
Thus, the peg was one of the
causes of the bubble, the other being that banks
were allowed to offer ``unhealthy'' loan conditions.
\qpar

As documented in the notes of Table A1
the decision of September 1983 to peg 
the Hong Kong currency to the USD 
under a currency board system did not meet the
approval of the people of
Hong Kong.

\begin{table}[htb]

\small

\centerline{\bf Table A1\quad Chronology of the 
monetary system of Hong Kong}

\vskip 5mm
\hrule
\vskip 0.7mm
\hrule
\vskip 2mm

$$ \matrix{
\qtb
\hbox{}&1895-1935 & 1935-1941 & 1941-1945 & 1945-1972 & 1972-1974 & 
1974-1983& 1983-? \cr
\noalign{\hrule}
\qth
\hbox{Peg}&\hbox{British} &\hbox{British} &\hbox{Japanese}
&\hbox{British} &\hbox{USD} & \hbox{No peg} &\hbox{USD} \cr
\hbox{}&\hbox{pound} &\hbox{pound} &\hbox{yen} &\hbox{pound} &\hbox{} &
\hbox{} &\hbox{}  \cr
\qtb
\hbox{CB}&\hbox{yes} &\hbox{yes} &\hbox{yes} &\hbox{yes} &
\hbox{\large ?} & \hbox{No} &\hbox{yes} \cr
\noalign{\hrule}
} $$
\vskip 1.5mm
\small
Notes: CB means ``currency board''. From 1895 to 1935
the currency of Hong Kong (as well as Singapore)
consisted of  ``British trade dollars'' minted in India.
During the British colonial time one can say that it was a
CB system in the sense that 
ultimately the monetary policy was decided in London.
The question mark indicates that so far we could not find
sufficient information about this episode.
Incidentally, it is often said that the decision to establish a
currency board was taken in the wake of the
panic following the ``Black Saturday''
of 24 Sep. 1983 when the exchange rate of the 
HKD fell to an all-time
low of USD 1 = HKD 9.6 (i.e. 23\% below its current rate).
In fact, according to John Greenwood (2008)
who played a key role in this story,
the draft of the project was written in early Sep. 1987, then
discussed with Milton Friedman and Maxwell Fry (professor
at the University of Hawaii).Then, the project  was sent
to Washington where it was approved by the British PM
(probably after also consulting with President Reagan)
at a meeting
held at the British Embassy in Washington on Tuesday 27 Sep.
1987. At the beginning of his account, Greenwood writes that
``there was solid opposition in Hong Kong to my proposal
for the restoration of a currency board''. In a sense
this is understandable because it meant upholding the
colonial regime. \qL
{\it Sources: Pershing Square (2011), Greenwood (2008).
The second source is not completely reliable for it says
that ``following the Thatcher visit to Beijing in September 1982,
the HK dollar had 
began to depreciate until in mid-September 1983 inflation 
surged to 18\%''. In fact the inflation rate was 16\% in Sep. 1981,
then dropped steadily to 9\% in Sep. 1982. In Sep. 1983 inflation
stood around 11\%.}
\vskip 5mm
\hrule
\vskip 0.7mm
\hrule
\end{table}

A peg can work without currency board. As an illustration
one may consider the case of Saudi Arabia. 
Although Saudi Arabia does not have a currency board
system its currency, the riyal,
is pegged to the USD (currently 1 USD=3.75 riyal) since
1960. Until 1986 the fixed exchange rate was adjusted
periodically (with changes of less than 2\% usually) but since
1986 the exchange rate has remained unchanged. Even
in this system without currency board 
there is nevertheless a mandatory 100\% currency backing of riyals
emission by foreign exchange reserves (Al-Jasser 2005, p. 265 and 270).
Thus, the main difference
with Hong Kong is that in Hong Kong the foreign exchange reserve
{\it must} be in USD whereas in Saudi Arabia it can be in any
reserve currency. As Saudi Arabia's oil income is completely in USD
in practice there is little difference between the two cases.
\qpar

Why did we give a close attention to the question of foreign
exchange reserves in Hong Kong? The answer is simple.
Within a few years, as the real
estate crises develops, this may become a key issue for Hong
Kong. It may emerge and unfold through the following steps.
\qee{1} First of all, it should be observed that for any currency
(as indeed for any commodity) a high demand is better than a
sluggish or declining demand. The reason is obvious. Historically,
many sovereign debt crises were triggered by falling exchange
rates but, to our best knowledge, none was triggered
by a rising exchange rate%
\qfoot{It is true that a strengthening
exchange rate may adversely affect the balance of trade, but the
case of the United States over past decades shows that a major
country is not much affected by a negative balance of payments.}%
.
\qee{2}
Therefore for any country, and especially if its
currency is a reserve currency,
all situations which increase the demand for that currency
go in the right direction, e.g. when  USD are required to buy oil
or other commodities or when the issuance of a country's currency
must be backed by USD (as is the case of any country 
which has a currency board based on the USD). In the early 1990s
US economist Steve Hanke, a former Chief Economic Adviser
to President Reagan,
visited South American and
East European countries
to give advice to their governments on the currency board system.
His attempts met temporary success in Argentina, Bulgaria, Estonia,
Lithuania and Bosnia-Herzegovina (for more details see
Hanke's biographical article on Wikipedia).
However, Argentina dropped
the dollar peg during the crisis of 2000-2002 while the four
European countries eventually pegged their currencies to the euro.
\qee{3} In two or three years 
when property prices will have fallen by 50\% or more
it is likely that some persons, both in Hong Kong and Beijing,
will observe that the crisis was engineered by the peg to
the dollar and will suggest that
it be changed to a peg to the renminbi
or at least to a basket of currencies including the renminbi.
Actually, such a move has already been predicted and anticipated
by some hedge funds (see Pershing Square 2011). 

\qpar
There is no conflict of interest.

\vskip 5mm

{\bf References}

\qparr
Al-Jasser (M.), Banafe (A.) 2005: Foreign exchange intervention
in Saudi Arabia. 
Bank of International Settlements Papers No 24, 265-272.

\qparr
Bradsher (K.) 2002: New challenge for China's shaky banks.
New York Times, 17 September 2002.

\qparr
Bradsher (K.) 2003: Property slump ruins many in Hong Kong.
New York Times, 15 August 15 2003.

\qparr
Greenwood (J.) 2008: Hong Kong's link to the US dollar.
Origins and evolution.
Hong Kong University Press, Hong Kong.

\qparr
Holzhey (M.), Skocsek (M.) 2015: UBS Global Real Estate Bubble
Index for housing markets of selected world cities. 
UBS Switzerland AG (15 October 2015).

\qparr
Hong Kong Housing Authority 2015: Housing in figures.

\qparr
Hong Kong Monetary Authority 1994: The interest rate
structure in Hong Kong. Quarterly Bulletin, November 1994.

\qparr
Maslov (S.), Roehner (B.M.) 2003:
Does the price multiplier effect also hold for stocks?
International Journal of Modern Physics C 14,10,1439-1451.

\qparr
Pershing Square Capital Management 2011: Linked to win.
Published by the hedge fund ``Pershing Square'' on 11 September
2011.

\qparr
Richmond (P.) 2007: A roof over your head; house price peaks in the UK
and Ireland.
Physica A, 375,1,281-287.

\qparr
Richmond (P.), Roehner (B.M.) 2013: 
The predictable outcome of house price peaks. 
Evolutionary and Institutional Economic Review 9,1,125-139.

\qparr
Roehner (B.M.) 2000: Speculative trading: the price multiplier effect.
The European Physical Journal B 14,395-399.

\qparr
Roehner (B.M.) 2001: Hidden collective factors in speculative
trading. Springer-Verlag, Berlin. Second edition with four
new chapters (2009).

\qparr
Roehner (B.M.) 2002, 2004:
Patterns of speculation in real estate and stocks.
Nikkei Workshop, 11 November 2002. 
published in Takayasu, editor: The application of
econophysics (2004). Springer-Verlag, Tokyo.

\qparr
Roehner (B.M.) 2006: 
Real estate price peaks: a comparative perspective.
Evolutionary and Institutional Economics Review 2,2,167-182.

\end{document}